\input harvmac
\input epsf
%\draftmode
\def\figin{\epsfcheck\figin}\def\figins{\epsfcheck\figins}
\def\epsfcheck{\ifx\epsfbox\UnDeFiNeD
\message{(NO epsf.tex, FIGURES WILL BE IGNORED)}
\gdef\figin##1{\vskip2in}\gdef\figins##1{\hskip.5in}% blank space instead
\else\message{(FIGURES WILL BE INCLUDED)}%
\gdef\figin##1{##1}\gdef\figins##1{##1}\fi}
\def\DefWarn#1{}
\def\figinsert{\goodbreak\midinsert}
\def\ifig#1#2#3{\DefWarn#1\xdef#1{fig.~\the\figno}
\writedef{#1\leftbracket fig.\noexpand~\the\figno}%
\figinsert\figin{\centerline{#3}}\medskip\centerline{\vbox{\baselineskip12pt
\advance\hsize by -1truein\noindent\footnotefont{\bf Fig.~\the\figno:} #2}}
\bigskip\endinsert\global\advance\figno by1}

\def\npb{Nucl.Phys.}
\def\prd{Phys. Rev. }
\def\plb{Phys. Lett. }
\def\bb#1{ hep-th/#1}

\lref\vafa{C. Vafa, {\it Gas of D-Branes and Hagedorn Density of BPS States},
hep-th/9511088, Nucl.Phys. B463 (1996) 415-419.}

\lref\braneReview{D. Kutasov and E. Giveon, Rev.Mod.Phys. 71 (1999) 983-1084
, hep-th/9802067}
\lref\cw{S. Coleman and  E. Weinberg, Phys.Rev.D7:1888-1910,1973}
\lref\books{Books on string theory such as M. Green, J. Schwarz, and  E.
Witten
Cambridge, Uk: Univ. Pr. ( 1987) and
J. Polchinski, Cambridge, Uk: Univ. Pr. (1998)}
\lref\wb{J. Wess and J. Baggar, Princeton University Press 1992}
\lref\bss{T. Banks, N. Seiberg, and E. Silverstein,
\plb B401:30-37,1997; \bb9703052.}
\lref\bdg{C. Bachas, M. Douglas, and M. Green, JHEP 9707:002, 1997;
\bb9705074.}
\lref\kMc{S. Kachru and J. McGreevy, \prd D61:026001,2000; \bb9908135.}
\lref\bdl{M. Berkooz, M. Douglas, and R. Leigh, \npb B480:265-278, 1996;
\bb9606139.}
\lref\dkps{M. Douglas, D. Kabat, P. Pouliot, and S. Shenker
\npb B485:85-127,1997; \bb9608024.}
\lref\az{A. Hanany and A. Zaffaroni, \npb B509:145-168, 1998;
\bb9706047.}

\lref\bh{
J.~H.~Brodie and A.~Hanany,
``Type IIA superstrings, chiral symmetry, and N = 1 4D gauge theory  dualities,''
Nucl.\ Phys.\  {\bf B506}, 157 (1997)
[hep-th/9704043].}

\lref\romans{L.J. Romans, \plb B169:374, 1986.}
\lref\ght{M. Green, C. Hull, and P. Townsend,
\plb 382(1996)65-72;\bb9604119.}
\lref\strominger{A. Strominger, \plb B383:44-47, 1996; \bb9512059.}
\lref\ch{C.G.Callan and J.A. Harvey, \npb B250:427,1985.}
\lref\brodie{J.H.Brodie, \npb B532:137-152,1998; \bb9803140.}
\lref\CSreview{For a nice review see
G.V. Dunne, Les Houches Lectures 1998; \bb9902115.}
\lref\jl{J. Maldacena and L. Susskind, \npb B475:679-690,1996;
\bb9604042.}
\lref\redlich{Redlich, \prd D29(1984)2366.}
\lref\bt{J.H. Brodie and S. Thomas, to appear.}
\lref\zumino{See for example, B.~Zumino,
``Chiral Anomalies And Differential Geometry: Lectures Given At Les Houches,
August 1983,''
UCB-PTH-83/16
{\it Lectures given at Les Houches Summer School on Theoretical Physics, Les Houches, France, Aug 8 - Sep 2, 1983}.}
\lref\kaplan{D.~B.~Kaplan,
``A Method for simulating chiral fermions on the lattice,''
Phys.\ Lett.\  {\bf B288}, 342 (1992)
[hep-lat/9206013].}
\lref\witten{E.~Witten,
``Branes and the dynamics of {QCD},''
Nucl.\ Phys.\  {\bf B507}, 658 (1997)
[hep-th/9706109].}

\lref\skyrme{T.~H.~Skyrme, ``A Unified Field Theory Of Mesons And
Baryons,'' Nucl.\ Phys.\  {\bf 31} (1962) 556.}

%\cite{Seiberg:1994rs}
%\bibitem{Seiberg:1994rs}
\lref\SW{N.~Seiberg and E.~Witten, ``Electric - magnetic duality,
monopole condensation, and confinement in N=2 supersymmetric
Yang-Mills theory,'' Nucl.\ Phys.\ B {\bf 426}, 19 (1994)
[Erratum-ibid.\ B {\bf 430}, 485 (1994)] [arXiv:hep-th/9407087].}
%%CITATION = HEP-TH 9407087;%%

%\cite{Karch:1998yv}
%\bibitem{Karch:1998yv}
\lref\karch{A.~Karch, D.~Lust and D.~J.~Smith, ``Equivalence of
geometric engineering and Hanany-Witten via fractional  branes,''
Nucl.\ Phys.\ B {\bf 533}, 348 (1998) [arXiv:hep-th/9803232].}
%%CITATION = HEP-TH 9803232;%%

%\cite{Giveon:1998sr}
%\bibitem{Giveon:1998sr}
\lref\giveon{A.~Giveon and D.~Kutasov, ``Brane dynamics and gauge
theory,'' Rev.\ Mod.\ Phys.\  {\bf 71}, 983 (1999)
[arXiv:hep-th/9802067].}
%%CITATION = HEP-TH 9802067;%%

%\cite{Brodie:2001pw}
%\bibitem{Brodie:2001pw}
\lref\brodie{ J.~H.~Brodie,
%``On mediating supersymmetry breaking in D-brane models,''
arXiv:hep-th/0101115.}
%%CITATION = HEP-TH 0101115;%%

\lref\hooft{G. 't Hooft, Nucl.\ Phys. B{\bf75}461(1974)}

\def\and{&}

% ==========================================================================
% Title page
% ==========================================================================

\def\LongTitle#1#2#3#4#5{\nopagenumbers\abstractfont
\hsize=\hstitle\rightline{#1}
\hsize=\hstitle\rightline{#2}
\hsize=\hstitle\rightline{#3}
\vskip 0.5in\centerline{\titlefont #4} \centerline{\titlefont #5}
\abstractfont\vskip .3in\pageno=0}

\LongTitle{}{hep-th/0208191}{} {Vortices}{ under S-duality}

\centerline{
  John H. Brodie      }
\bigskip
\centerline{Perimeter Institute} \centerline{Waterloo, ON}
\centerline{jbrodie@perimeterinstitute.ca}\vskip 0.3in
\centerline{\bf Abstract} We investigate a system of 1,3, and
7-branes in type IIB string theory. We identify the 1-brane with
the vortex of the Abelian Higgs model. We claim that in the
S-dual, the theory on the 3-brane confines electric flux which we
identify with the fundamental string and that the dual photon has
a mass . We then investigate ways of breaking supersymmetry in the
brane set up.

\Date{August 2002}
\bigskip

\newsec{Introduction.}

Confinement in QCD is a mysterious phenomenon. Despite years of
effort analytically and on the lattice there is still no proof
that QCD confines. There are many heuristic arguments that QCD aw
well as pure Yang-Mills theory do confine. One argument is that at
large N the Wilson loop becomes a string world sheet \hooft.
Another argument relies on analogies with superconductivity. There
the electric photon gets a mass confining magnetic charge. In QCD
the dual photon or gluons supposedly gets a mass confining
electric charge. Although the duality works for abelian theories
generalizing it to non-Abelian theories has been difficult. In
1994 Seiberg and Witten \SW\ showed that certain  supersymmetric
theories confine in the IR. The point of this paper is to consider
a D-brane construction whose low energy limit is the Abelian-Higgs
model. By turning on a FI-term, we can have vortex solitons.
D1-branes play the role of the magnetic vortices. Since the brane
set-up is in type IIB we can S-dualize the brane configuration.
D1-strings become F1-strings. The dual theory is electrically
confining and posses electric vortices, the F1-strings.

\newsec{Abelian Higgs model}

Consider the Abelian Higgs model. This is a $U(1)$ gauge theory
coupled to a charged scalar field. In supersymmetric language the
Abelian gauge field gets a mass by eating the hypermultiplet. The
hypers get a vev from a D-term. In superspace notation the action
is

\eqn\ss{L = \int  d^2\theta {1\over g^2} W_{\alpha}^2 + \int
d^4\theta Q^{\dag}e^V Q + \tilde Q^{\dag} e^{-V} \tilde Q - \xi V
} In components the bosonic part of this action is

\eqn\comp{ L = {1\over g^2}F^2 + D_{\mu} q^*D^{\mu} q + (|q|^2 -
|\tilde q|^2 - \xi)^2}

 This model has BPS vortex solutions. The mass of a vortex in
three dimension or the tension of a string in four dimensions goes
as the Higgs vacuum expectation value $\xi$  divided by the gauge
theory coupling constant.

\eqn\mass{M_v^2 = \xi/g^2}

Supersymmetry breaking can be achieved by giving a mass to the
gauginos at tree level. This mass can be made of order the mass of
the Higgs field without destabilizing the Higgs potential. In this
way the problem of stability is similar to that of the standard
model with supersymmetry being the cure.

\newsec{brane set up}

\ifig\dsix{We consider the T-dual of this system. Here we see a
D6-brane tilted with a D4-brane splitting on it. The horizontal
direction is compact. The thick line is the D2-brane that connects
the two ends of D4-brane. The D2-brane has the interpretation as a
vortex in 3+1 dimensions.} {\epsfxsize2.0in\epsfbox{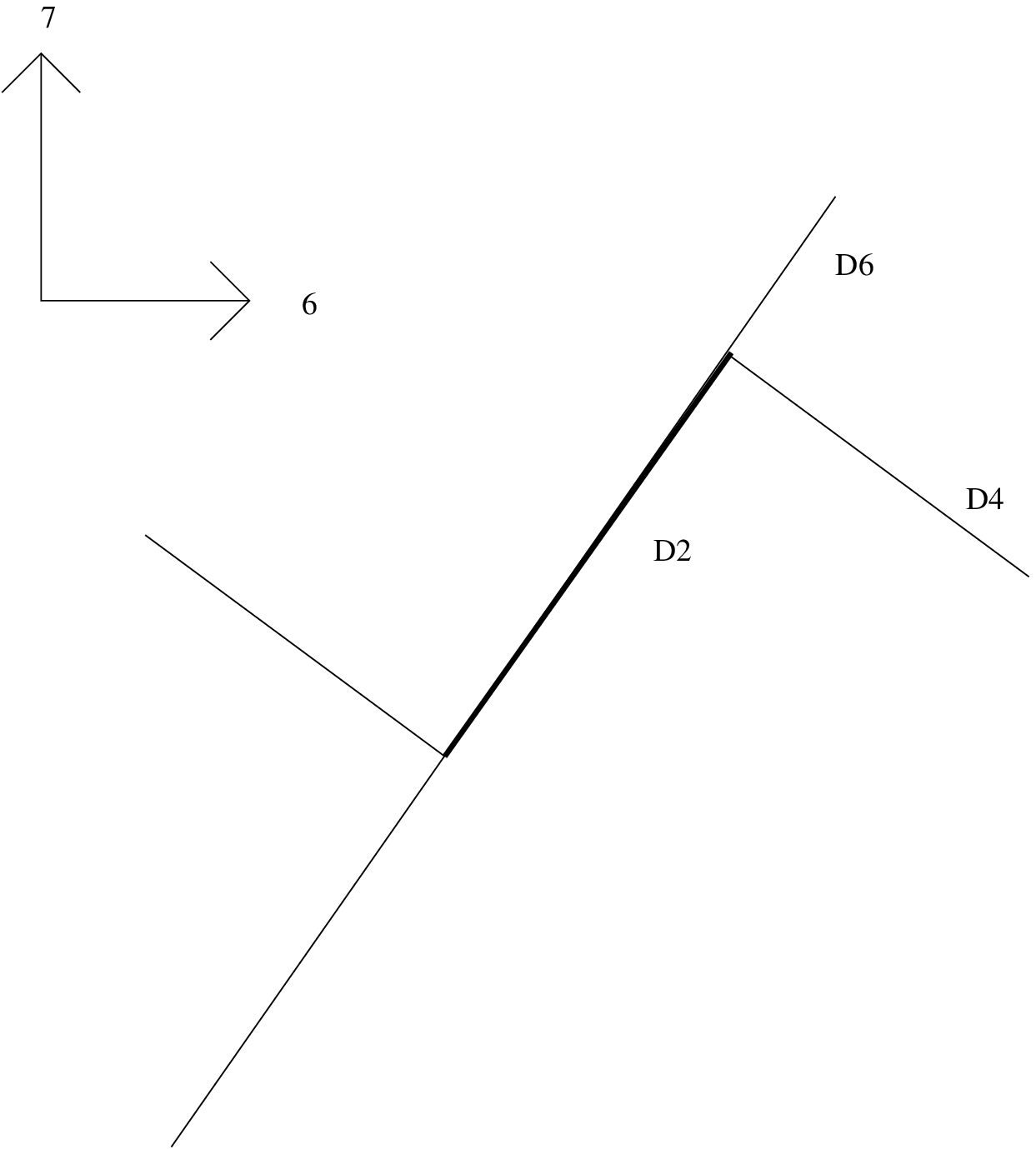}}

The Abelian Higgs model can be realized in string theory as a
3-brane in directions 123 parallel to a 7brane in directions
1234567. The FI term is proportional to NS-NS B field in the 4567
direction. A 1brane can then be bound to the 3 brane if a B field
is turned on in the 2345. This is easiest to understand in a
T-dual picture of 4-branes and 6branes perpendicular \giveon. The
B-field in the 67 direction along the D7-brane under T-duality
becomes proportional to the angle of rotation of the D6-brane
\brodie. The B-field in the 45 direction becomes rotation of the
4-brane after it splits on the 6-brane. Now there is in the T-dual
picture a 2-brane which is parallel to the 4-brane. However, the
4-brane has now rotated and split on the rotated 6-brane to
decrease its length. The 2-brane can now split on the 4-brane in
the same way that the 4-brane split on the 6-brane. In the 4+1
theory the 2-brane is a magnetic string dipole. As for magnetic
particle dipoles the region between the positive and negative
monopoles is a solenoid. The effect is the same here but in one
dimension higher. The interpretation of the splitting of the
2brane on the 4-brane for the 1+1 theory living on the
compactified 2-brane world volume is the same as it was for the
splitting of the D4-branes on the D6-brane: the Abelian gauge
field in 1+1 has Higgsed. The tilt of the 2-brane now in the
67-plane corresponds under T-duality to 3-brane charge in 0167.
This can be interpreted as B field  in the 2345 direction.

The  system of 1-3-7 branes in IIB is S dual to a 3brane bound to
a non perturbative 7brane with RR B field turned on. This theory
is the magnetic dual of the Abelian Higgs model. In this theory
the magnetic photon gets a mass proportional to the Higgs vev
divided by the gauge theory coupling constant, confining electric
charge. The electric confinement produces electric vortices.
Indeed the D1brane under S duality becomes a fundamental string
with tension proportional to the Higgs vev. These fundamental
strings do not spread out in the 3brane but remain confined
strings of electric flux. This should be of no surprise since this
brane configuration corresponds to the low energy limit of $N=2$
$Sp(1)$ theory . The O7-plane splits up into two non-perturbative
D7-branes corresponding to the monopole point and the dyon point
of \SW. Here we are considering Higgsing along the monopole point.

 Supersymmetry breaking can
be implemented in the S dual picture by a fractional brane
orbifold \karch. As was said the supersymmetry breaking scale
cannot be much greater than the higgs mass. In this case the dual
photon mass is much higher than the Higgs mass since its mass
comes with a gauge coupling constant squared in its denominator.
However, the string tension is proportional to the Higgs mass and
not the dual photon mass. Therefore the supersymmetry breaking
scale cannot be much larger than the confinement scale.

\newsec{2+1 vortices with less supersymmetry}
One can also consider D7-branes along directions 1234568,
D5-branes along 12347, D3's along 127, and D1-branes along the
7-direction. By rotating the D7-brane the D5 splits on the
D7brane, then the D3-brane splits on the D5, while the D1-brane
splits on the D3-brane. Now T-dual this configuration. The
D7-brane becomes an 8-brane with flux turned on $B_{ 78}$. The
D5-brane becomes a D4-brane in 1234 with flux turned on in 56. The
flux $B_{78}B_{56}$ corresponds to the D-term in the field theory.
Then T-dual of the 3-brane becomes a 2-brane 12 with flux turned
on in 3456. The 1-brane then becomes a 0-brane with flux turned on
in 123456. The low energy effective theory now is a 2+1 gauge
theory with vortices.

\newsec{Acknowledgements}
We would like to thank E. Poppitz, M. Strassler, and D. Tong for
helpful discussions.

\listrefs

\end